# A Modified Approximate Dynamic Programming Algorithm for Community-level Food Security Following Disasters

Saeed Nozhati[a], Yugandhar Sarkale[b], Bruce R. Ellingwood[a], Edwin K.P. Chong[b], Hussam Mahmoud[a]

[a] Department of Civil and Environmental Engineering, Colorado State University, Fort Collins, CO 80523-1372, USA  Saeed.Nozhati, Bruce.Ellingwood, Hussam.Mahmoud@colostate.edu

[b] Department of Electrical and Computer Engineering, Colorado State University, Fort Collins, CO 80523-1373, USA  Yugandhar.Sarkale, Edwin.Chong@colostate.edu

**Abstract:**  In the aftermath of an extreme natural hazard, community residents must have access to functioning food retailers to maintain food security. Food security is dependent on supporting critical infrastructure systems, including electricity, potable water, and transportation. An understanding of the response of such interdependent networks and the process of post-disaster recovery is the cornerstone of an efficient emergency management plan. In this study, the interconnectedness among different critical facilities, such as electrical power networks, water networks, highway bridges, and food retailers, is modeled. The study considers various sources of uncertainty and complexity in the recovery process of a community to capture the stochastic behavior of the spatially distributed infrastructure systems. The study utilizes an approximate dynamic programming (ADP) framework to allocate resources to restore infrastructure components efficiently. The proposed ADP scheme enables us to identify near-optimal restoration decisions at the community level. Furthermore, we employ a simulated annealing (SA) algorithm to complement the proposed ADP framework and to identify near-optimal actions accurately. In the sequel, we use the City of Gilroy, California, USA to illustrate the applicability of the proposed methodology following a severe earthquake. The approach can be implemented efficiently to identify practical policy interventions to hasten recovery of food systems and to reduce adverse food-insecurity impacts for other hazards and communities.

1. **INTRODUCTION**

A resilient food supply is necessary for securing and maintaining an adequate food stock for urban inhabitants before and following extreme hazard events, such as earthquakes and hurricanes. Food security issues are exacerbated during the chaotic circumstances following disruptive hazard events Biehl et al. (2017). Such events can damage food systems, household units, and utility and transportation systems, thereby endangering public health and community food security. Main food retailers play a pivotal role in ensuring community food security, and their functionality, along with the functionality of household units, is a critical concern of community leaders. Even though numerous efforts have been undertaken to mitigate the challenging problem of food shortage, very few studies address this problem from a systems perspective where the interaction between various network is considered simultaneously. We propose a holistic approach that not only considers the interactions between the critical networks that contribute towards food security but also provide a method to compute the near-optimal recovery actions at the community level following the occurrence of extreme natural hazards. The proposed decision-making algorithm can handle large-scale networks and provides robust and anticipatory decisions. To this end, we leverage the approximate dynamic programming (ADP) paradigm to calculate the near-optimal actions periodically and employ a simulated annealing algorithm to guide the ADP method in selecting the most promising recovery actions (see Bertsekas (1995) for a detailed description of ADP). This fusion enables us to calculate the recovery actions in a limited time, which is usually not possible because of large candidate solutions. A testbed community modeled after Gilroy, California, is presented to illustrate how the proposed approach can be implemented efficiently to find the optimal decisions. Our approach provides policies to restore the critical Electrical Power Networks (EPN), Water Networks (WN), and highway bridges of Gilroy, in the aftermath of a severe earthquake, in a timely fashion. Specifically, we find a near-optimal sequence of decisions such that the main utilities of electricity and potable water are restored to the main food retailers and household units in (approximately) the shortest amount of time. Additionally, we also make sure that functional food retailers are accessible to community residents. This study pursues the Sustainable Development Goals (SDGs), a collection of 17 global goals set by the United Nations. For example, a result of this study is trying to make communities and human settlements inclusive, safe, resilient and sustainable (SDG 11). We aim to guide community leaders and risk-informed decision makers in reducing adverse food-insecurity impacts from extreme natural hazards.

2. **PRELIMINARIES**

In this section, we describe the fundamental methods of approximate dynamic programming and simulated annealing. The fusion of these two methods to obtain near-optimal recovery actions for our problem is illustrated in our case study in Section 4.

**2.1. Approximate Dynamic Programming**

In theory, dynamic programming (DP) provides an exact computable solution to a finite horizon scheduling problem. However, for many realistic scheduling problems, an exact solution of the problem by DP is practically impossible owing to several pragmatic constraints: for one thing, there is an exponential increase in the number of computations as the problem size grows; for another, DP computations do not usually accommodate time constraints into the solution process. Furthermore, many problems require *on-line replanning*, where the solutions must be adaptable to the availability of new data Bertsekas (1995).

A detailed discussion on the dynamic programming approach is beyond the scope of this text, and we generously adopt the notation from our previous studies; therefore, the interested reader is advised to refer to the study by Nozhati et al. (2018a) for a detailed understanding of the notation employed here. Consider the minimization of the objective function $F(x_1, \ldots, x_N)$ of the discrete variables $x_i$ that takes values in a finite set. This optimization problem can be solved using various approaches. In dynamic programming, we have

$$x_\alpha^* \in \arg\min_{x_\alpha} J_\alpha(x_1^*,...,x_{\alpha-1}^*, x_\alpha) \tag{1}$$

$$J_\alpha(x_1, x_2,..., x_k, x_\alpha) = \min_{x_{k+1},...,x_N} F(x_1,...,x_k, x_{k+1},...,x_N) \tag{2}$$

where $x_\alpha^*$ is the optimal value of $x_\alpha$ and $J_k$ are called the optimal cost-to-go functions, defined by the recursion

$$J_k(x_1, x_2,..., x_k) = \min_{x_{k+1}} J_{k+1}(x_1, x_2,..., x_k, x_{k+1}) \tag{3}$$

Unfortunately, computing such cost-to-go functions is virtually impossible in many real-world problems. Hence, dynamic programming approaches must employ approximations of the cost-to-go functions. One specific cost-to-go approximation is *rollout*. In the rollout methods, $J_k$ are approximated by a base heuristic ($H$), and the resultant approximation is denoted by $H_k(x_1, ..., x_k)$. Thus, the *rollout* algorithm calculates a solution by replacing $J_k$ with $H_k$ in (3):

$$\tilde{x}_k \in \arg\min_{x_k} H_k(\tilde{x}_1,...,\tilde{x}_{k-1}, x_k) \tag{4}$$

Typically, the base heuristic is defined to be the recovery policies of responsible public and private entities. For a more detailed description of the base heuristic, see Nozhati et al. (2018a). Rollout methods evaluate the performance of $H$ and calculate an improved plan (*rollout policy*) by using a *single-step lookahead*. A multi-step lookahead is also possible, but at significantly increased computational cost. The efficiency of the method is appealing, especially for discrete and deterministic optimization problems. An important and attractive feature of this method is the improved performance that it yields over the underlying base heuristic. This improvement often is remarkable in practical cases.

## 2.2. Simulated Annealing

Simulated annealing (SA) is a random-search technique for global optimization problems. Deterministic search and gradient-based methods often get trapped at local optima during the search process. However, the SA method overcomes this limitation and converges to a globally optimal solution, provided enough iterations are performed and a sufficiently slow cooling schedule is employed. The underlying Markov chain in SA not only accepts changes that improve the objective function but also keeps some changes that are not ideal Yang (2010). The likelihood of acceptance of a worse solution decreases with the difference in objective function values. More precisely, the probability of acceptance of a worse solution is given by

$$P = e^{-\frac{\Delta f}{k_B T}} \tag{5}$$

where $\Delta f$ is the change in objective function, $T$ is a positive real number representing the current temperature, and $k_B$ is Boltzmann's constant. The search process would be a greedy search provided that $T \to 0, P \to 0$, because in this case it only accepts better solutions. On the other hand, the search process would be a random selection process provided that $T \to \infty, P \to 1$, because here it accepts any solution.

## 3. CASE STUDY

Gilroy is a city of approximately 50,000 located in Northern California's Santa Clara County, which is approximately 12 km from the San Andreas Fault, and is used to illustrate the proposed method. The availability of reasonable information about EPN, WN, population density, main food retailers, and the high level of seismic exposure of the area influenced us in adopting Gilroy as our case study. To perform risk and resilience assessment at the community level, we consider the electrical power network (EPN), potable water network (WN), main food retailers, and highway bridges. In this section, we briefly describe the features of mentioned infrastructure systems; the reader interested in a more thorough treatment of modeling and description of EPN and WN should refer to Nozhati et al. (2018a, 2018b) and Sarkale et al. (2018).

### 3.1. Electrical Power and Water Networks

The EPN plays a crucial role in the functionality and well-being of a community in that the operation of other networks is heavily dependent on the functionality of the EPN. EPN malfunction can disrupt the functionality of critical facilities like main food retailers. The modeled EPN of Gilroy within the defined boundary is shown in Figure 1a. Steel lattice towers supply the Llagas substation, the main source of power. The subtransmission towers that support electrical lines that supply main food retailers, water pumps, and urban grids are modeled and spaced at 100 m.

The functionality of the potable water network is of great significance, especially following disasters, to support inhabitants' health, firefighting, and industrial processes. The major components of the WN in Gilroy are illustrated in Figure 1b. The WN consists of six water wells, two booster pump stations (BPS), three water tanks (WT), and the main pipelines.

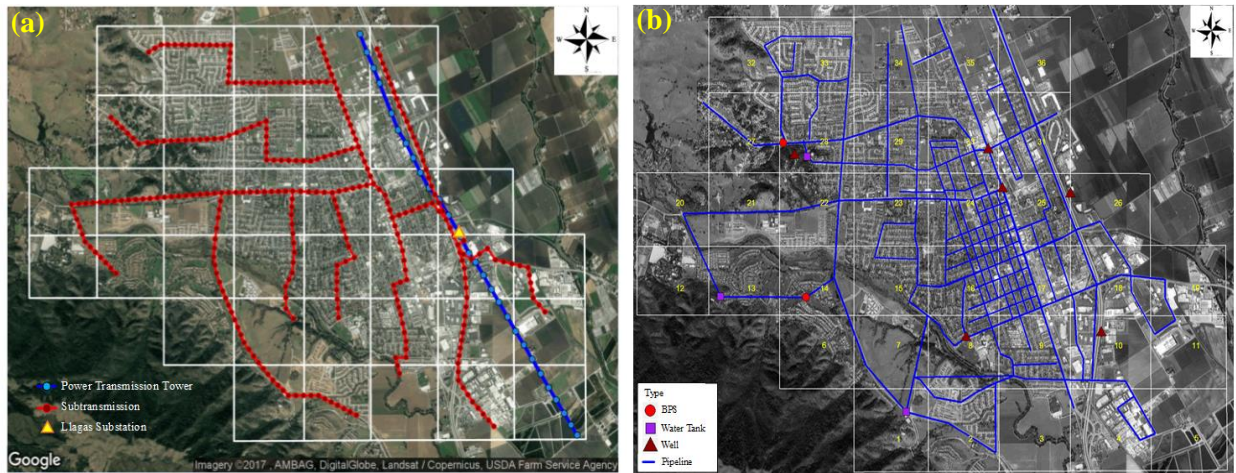

Figure 1. a) Electrical Power Network b) Water Network

### 3.2. Food Retailers

Six main food retailers—Walmart, Costco, Nob Hill Foods, Safeway, and Target—serve the community of Gilroy. The availability of electricity and potable water is crucial for the functioning of these retailers and for the food security of the community. To determine the shopping-activity probabilities of Gilroy inhabitants based on the main retailers, we use the gravity model proposed in the study Adigaa et al. (2014). The probabilities calculated using the gravity model imply that people are more likely to shop at closer and bigger retailers, which indicates that transportation accessibility is an important factor in community food security and is discussed next.

### 3.3. Highway Bridges

The critical facilities and residence units not only must have essential utilities but also should be accessible. Gilroy is at the intersection of two main highways: U.S. 101, which extends through the City in a north/south route, and SR 152, which extends in an east/west direction. Several highway bridges must be operable to serve the transportation of the community and accessibility, especially for the main food retailers of Costco, Walmart, and Target.

### 3.4. Seismic Hazard Simulation

We consider a scenario earthquake similar to the Loma Prieta Earthquake of 1989 ($M_w$ = 6.9), the most devastating earthquake that Gilroy has experienced in recent times. This seismic event has a hypothetical epicenter set roughly 12 km southwest of downtown Gilroy on a projection of the San Andreas Fault. We use the Abrahamson et al. (2013) ground motion prediction equation (GMPE) to compute the effect of this simulated event on the components of the community. We model each component of infrastructure systems as a point within the topology of the community; the *fragility curves* and *restoration times* based on the level of damage are assigned to the components to evaluate their seismic performance following the earthquake. For further information regarding the seismic model and the restoration times, see Nozhati et al. (2018a, 2018b), Sarkale et al. (2018), Masoomi and van de Lindt (2018), and Masoomi et al. (2018).

## 4. POLICY OPTIMIZATION FOR FOOD SECUTIY

We illustrate the performance of the method using a customary objective in community resilience: the number of people whose residence units have electricity and potable water (*main utilities* for short), and who have *access* to a food retailer that also has service from the main utilities. A benefit to people characterized in this fashion captures the availability and accessibility of food retailers in our objective function. Our goal ($F$) is to compute the repair actions ($X$) to maximize the number of benefited people in the shortest possible time. Policymakers always face the constraints of limited resources in terms of available tools and repair crews, denoted as $N$ in this study. Suppose that the recovery decisions are performed at discrete times denoted by $t$. Let $D_t$ be the set of all damaged components before a repair action $x_t$ is performed, and $P_N(D_t)$ is the power set order $N$ of $D_t$ (see Nozhati et al. (2018a)). When all the components are repaired, $X = (x_1, \ldots, x_{t_{end}})$ is the set of repair actions. Let $k_t$ denote the total time elapsed until the completion of all repair actions, and $h_t$ be the total number of benefited people because of the repair action $x_t$, where $x_t \in P_N(D_t)$. Therefore, the objective function and the optimal solution $X^*$ are given by:

$$F(X) = \frac{1}{k_{t_{end}}} \sum_{t=1}^{t_{end}} h_t \times k_t \qquad (6a)$$

$$X^* := \arg\max_X F(X) \qquad (6b)$$

The $|P_N(D_t)|$ at each decision time $t$ is very large for community-level planning, especially when several networks are considered simultaneously and during initial stages of decision making. Therefore, at each decision time $t$, we employ the SA algorithm to search in the set $P_N(\tilde{D}_t)$, where $|\tilde{D}_t| < |D_t|$, $\tilde{D}_t$ is a subset of $D_t$, and to avoid searching over the entire set $P_N(D_t)$ exhaustively. A fixed number of iterations are provided to the simulated annealing algorithm. At each such iteration, $P_N(\tilde{D}_t)$ is recalculated to eliminate unpromising actions and incorporate new actions that are not considered in the previous iterations according to the probability of acceptance of worse solution mentioned in Section 2.2. Such a restriction on the number of iterations captures the constraints on the amount of time to be expended in calculating the optimal recovery actions at every $t$ and the solution accuracy warranted of each candidate recovery action at $t$. Despite this restriction, we show in the simulation results that the combined approach significantly improves over the recovery actions calculated using $H$.

## 5. RESULTS AND DISCUSSIONS

Once we subject Gilroy, California to the simulated earthquake, we calculate the damage to the individual components and initiate the recovery process. In this study, $H$ is chosen to be a random base restoration policy without any pre-assumption to show the effectiveness of the proposed method. Thereafter, the recovery actions are calculated using the fused method explained previously.

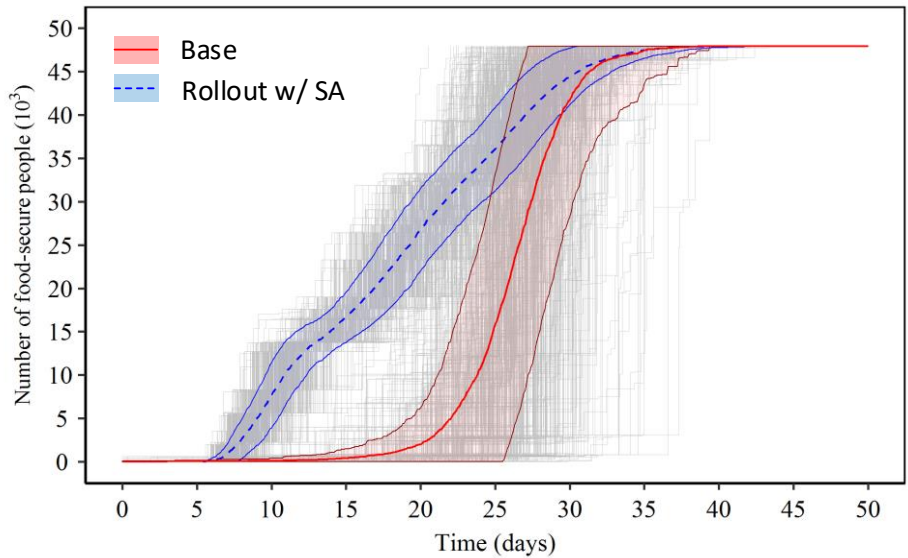

Figure 2. Comparison of base and rollout with simulated annealing policies

The rollout algorithm sequentially and consistently improves the underlying *H,* and the SA algorithm guides the rollout search to find the near-optimal actions at each stage non-exhaustively. This non-exhaustive guidance helps in limiting the amount of computations that results in increased solution speeds without affecting the performance of the rollout approach.

Figure 2 shows the performance of the proposed fused algorithm in the restoration of the defined community. Recall that the number of food-secure people are defined to be people who have functional main utilities and have accessible and available food retailers. The colored plots presented here are average (and one-standard-deviation band) of recovery for multiple damage scenarios, and the recovery in each such damage scenario is represented by the faint lines in Figure 2. Note how the rollout with simulated annealing (Rollout w/ SA) policy results in recovery actions that benefit a larger number of people per-time than the underlying base heuristic (Base). This is validated by calculating the area under the average plots (represented by the dotted blue and dark red lines) owing to Rollout w/ SA and Base. The area under the curve measures the benefit or impact of the recovery actions. The larger the area, the bigger the impact/benefit (normalized by the total recovery time). In the plots depicted in the Figure 2, it is obvious that the area under the curve of recovery owing to rollout w/ SA (average curve) is greater than the area under the recovery owing to Base (average curve). However, such a drastic improvement might not be realistic in all the cases. We attribute such a stark improvement in performance of the Rollout w/ SA algorithm to the fact that the performance of Rollout w/ SA is validated by comparing it with the performance of a random base heuristic, which does not offer a good performance in itself. Nonetheless, note that our algorithm utilizes this random base heuristic and improves its performance drastically; therefore, a comparison with the Base performance is justified. When other types of base heuristics are considered, both the average plots (Base and Rollout w/ SA) might intersect each other at multiple places, like in the study of Nozhati et al. (2018a). In all such scenarios, instead of focusing on individual sections of the average recovery, we must calculate the area under the recovery owing to Rollout and compare this calculated value with area under the recovery owing to Base (normalized by the total recovery time). This is in accordance with the definition of our optimization objective function as in (6a). Similarly, the one-standard deviation plots might intersect with each other; the area under the upper one-standard-deviation curve owing to Rollout w/ SA must be compared to the upper one-standard-deviation curve owing to Base. This is further illustrated in the Figure 3.

Figure 3 demonstrates the efficiency of the proposed methodology in terms of the final computed rewards, where reward is as defined in (6a). Such a plot provides an alternative perspective on the interpretation of the results. These rewards are calculated for multiple scenarios and are the area under the curve divided by the total recovery time for each of the faint plots depicted in the figure 2. The computed near-optimal repair actions using the Rollout w/ SA approach result in greater rewards than Base. This is validated by the separation between the two histograms shown in Figure 3. When the underlying base heuristic is no longer random, the two plots will intersect with each other; therefore, the performance evaluation because of the separation between the two histograms might not be apparent. In this situation, we extend the benefit analogy described for Figure 2. Particularly, we can calculate the benefit by using Figure 3 as follows: Count the number of samples in each bin for recovery owing to Base. Multiply the samples and the rewards for that bin. Repeat this for all the bins. Sum all these values to get the total impact owing to Base. A similar procedure can be followed for Rollout w/ SA and the two calculated values for Rollout w/ SA and Base can be compared to evaluate the performance of our method. As discussed previously, for the simulation results in Figure 2 and Figure 3, the benefit owing to Rollout w/ SA is more than that owing to Base; as can be seen by the separation between the maximum bin of Base histogram and the minimum bin of Rollout w/ SA histogram along the X-axis of Figure 3 (given that the total sum of samples, which is the number of scenarios is same for both the histograms). Therefore, in the Figure 3, we omit plotting the sample count along the Y-axis.

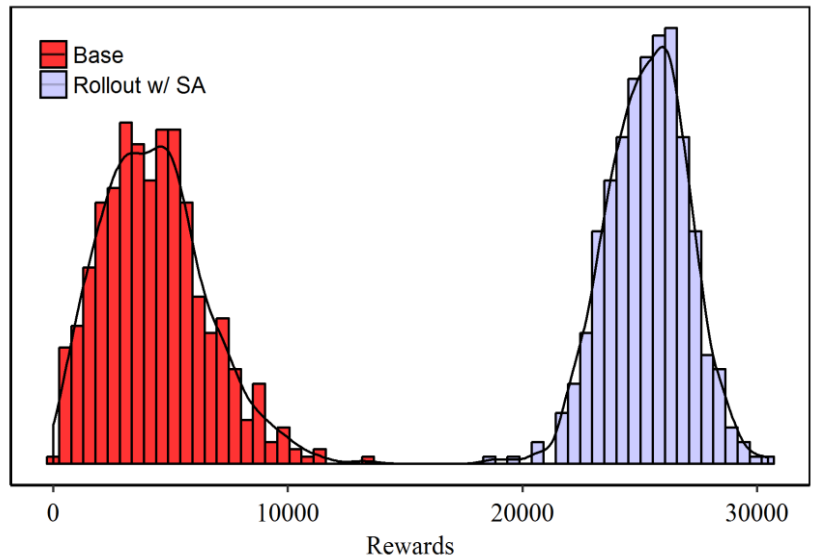

Figure 3. Histogram of F(X) for base and rollout with simulated annealing policies

## 6. CONCLUSIONS AND FUTURE WORK

We proposed an optimization formulation based on the approximate dynamic programming approach fused with the simulated annealing algorithm to determine near-optimal recovery policies in urban communities. To show the efficiency of the applicability of the method on the food security of realistic communities, the community of Gilroy (California) was modeled. Because the availability of utilities for people and retailers degrades the food security issues following a severe natural hazard event, the methodology optimizes the recovery of EPN, WN, and highway bridges so that a maximum number of people and main food retailers benefit from the prompt restoration of the main utilities.

For future work, we will incorporate the impact of structural and non-structural performance and contents damage to residential units and food retailers' establishments. Therefore, the functionality of household units and food retailers would contain the availability of the main utilities and building performance. While in this study we considered the availability and accessibility of food retailers, in a separate ongoing study we are examining the element of affordability of food, thus introducing an additional dimension regarding the impact of natural hazard events on economically vulnerable segments of the community. Furthermore, we will also fuse other meta-heuristics with the ADP formulation, and evaluate their performance with respect to simulated annealing Kaveh and Soleimani (2015). Even though the methodology demonstrated here does not consider stochastic actions, we are further extending this work to demonstrate that our approach can successfully deal with the uncertainty in the outcome of the recovery actions and should not be considered as a limitation.

**ACKNOWELDGMENT**

This work was supported by the National Science Foundation under Grant CMMI-1638284. This support is gratefully acknowledged. Any opinions, findings, conclusions, or recommendations presented in this material are solely those of the authors and do not necessarily reflect the views of the National Science Foundation.